\documentclass{article}
\usepackage{graphicx}
\usepackage{cite}
\def\bq{\begin{eqnarray}}
\def\eq{\end{eqnarray}}

\begin{document}

\title{\bf Excitation of an inertial  Unruh detector in the Minkowski vacuum: a numerical
calculation using spherical modes}

\author {Nistor Nicolaevici
\\
\it Department of Physics, West University of Timi\c soara,
\\
\it V. P\^arvan 4, 300223, Timi\c soara, Romania}

\maketitle

\begin{abstract}

We consider the excitation of a finite-length inertial Unruh detector in the Minkowski
vacuum with an adiabatic switch on of the interaction in the infinite past and a sudden switch off
at finite times, and obtain the excitation probability via a numerical calculation using the
expansion of the quantum field in spherical modes. We evaluate first the excitation probabilities
for the final states of the field with one particle per mode, and then we sum over the modes. An
interesting feature is that, despite of the inertial trajectory and of the vacuum state of the field,
the multipole components of the excitation probability are time-dependent quantities. We make clear
how the multipole sum yields the time-independent probability characteristic to an inertial
trajectory. In passing, we point out that the excitation probability for a sudden switch on of the
interaction  in the infinite past is precisely twice as large as that for an adiabatic switch on.
The procedure can be easily extended to obtain the response  of the detector along radial
trajectories in spherically symmetric spacetimes.

\end{abstract}

\\
\section*{1. Introduction}

The usual way to obtain the response function of the Unruh detector \cite{birr, cris} in a curved
spacetime is based on an integral over the Wightman function of the field evaluated along the trajectory
of the detector. Unfortunately, analytic expressions for the Wightman function are available only for a
few number of spacetimes. The more frequent situation is when one can obtain a complete set of the modes
of the quantum\, field. In such a case, the calculation may be carried out by evaluating first the individual
excitation probabilities for the final states of the field  with one particle in each mode,\footnote{We
have in mind the usual result in the first order of perturbation theory \cite{birr}.} and as a
second step summing over the modes.

The intention of this paper is to present such a calculation, considering the simple problem
of an inertial detector in the Minkowski vacuum. The less trivial point is that we will use the $spherical$
modes of the field. Primarily, we are interested in this exercise as a first step towards a numerical
procedure for obtaining the detector's response in a spherically symmetric curved spacetime, when the
Wightman function of the field is not at hand. We believe that some of the features of the excitation
probabilities uncovered here could be useful in organizing the analogous calculation in a curved
background. For simplicity, we will focus on a massless scalar field, using the standard definition
of the Unruh detector \cite{birr}.

The response of an Unruh detector in the Minkowski vacuum along various trajectories was studied
in numerous papers (see e.g. Refs.~\cite{svai, pere, ramk, satz, schl, louk, obad, barb, nico, koth1, lin,
koth, hu, douk, abdo, bren, gara2, iso}). The well-known conclusion concerning inertial trajectories is that
the excitation rate vanishes. However, the net excitation probability generally does not vanish, which is an
inevitable consequence of the perturbation  due to the coupling to the field. The main aim of our paper is
to (1) evaluate the excitation probability for an inertial trajectory following the procedure using the
spherical modes of the field mentioned above, and (2) compare the result with the probability obtained via
the standard calculation based on the Wightman function. We will encounter integrals which seem not to allow
an analytic evaluation, so in the end we will rely on numerical calculations. The basic conclusion is that,
as expected, the two procedures lead to the same result.

Let us briefly mention the key points in our calculation. It is also well-known that the response of the
detector depends on the switch on and off of the interaction. We will consider here the case when the interaction
is $(a)$ adiabatically switched on in the infinite past and $(b)$ suddenly switched off at some finite
time. We recall that, for a sudden switch on or off of the interaction, the excitation probability evaluated
in the limit of a vanishing regulator of the Wightman function diverges \cite{svai, pere, ramk, satz}.
We will therefore have to keep the regulator finite. We will use the standard $i\varepsilon$ regularization,
obtained by summing over the modes with the convergence factor $e^{-\varepsilon \omega}$.

An important mention regarding the regularization of the Wightman function is the following. It was noted
some years ago by Schlicht \cite{schl} that numerical calculations using the $i\varepsilon$ regularization\,
do not lead to the correct excitation rates in the well-studied case of a uniformly accelerated detector.
The solution adopted in Ref.~\cite{schl} is a new regularization procedure, based on assuming a (fixed)
finite length of the detector in its proper frame.\footnote{This is essentially equivalent to introducing a
frequency cutoff in the detector's frame, which is to be contrasted to the fixed cutoff in the static Minkowski
frame in the standard $i\varepsilon$ regularization.} The new regularization yields\, a covariant regularized Wightman
function, it  leads to physically sensible rates for a large class of accelerated
trajectories \cite{louk, obad, barb, nico}, while for a static detector it coincides with the usual $i\varepsilon$
regularization. In addition, it provides a simple interpretation of the regulator $\varepsilon$ in terms
of the  proper length of the detector, which can thus be kept as a significant parameter in the theory.

Despite of these positive aspects,  we will stick here with the old\, $i\varepsilon$ regularization.
One reason is that with this choice we avoid the sophistication due to the interplay between the
regulator and the trajectory \cite{schl}, simplifying the form of the excitation
amplitudes. Another reason is that, as a consequence of the underlying Lorentz invariance of the theory,
for the inertial trajectories the covariant regularization \cite{schl} completely eliminates the dependence
on the velocity of the detector. It will be of some interest to see how the excitation probability
based on the noncovariant regularization depends on this parameter.

We should emphasize that the situation considered here is not very different from that in Ref.~\cite{svai},
where the interaction between the detector and the field is suddenly decoupled both in the future
$and$ in the past. However, as a consequence of the adiabatic coupling in the past, we
will obtain a different result. In fact, we will find that the excitation probability is precisely
twice\, as large as that in Ref.~\cite{svai} for an infinite interaction time (i.e. for the interaction
switched on in the infinite past and/or switched off in the infinite future). It seems that this
fact remained unnoticed in the literature.

Perhaps the most interesting aspect in our calculation is that the\, excitation probabilities\, for the
final states of\, the field with\, a given orbital quantum number $\ell$ are $time$-$dependent$ quantities.
It will be then a good question to see how these probabilities add up to yield the time-independent excitation
probability characteristic to an inertial trajectory of the detector. We will illustrate this mechanism with
a series of plots. As can be guessed, the answer lies in the infinite sum over $\ell$. It is this  mechanism
that we believe will still operate for radial trajectories in a spherically symmetric curved background, and
which can be used to simplify the calculations.

The paper is organized as follows. In the next section we recall some basic facts about the Unruh detectors. In
Sec.~3 we obtain the excitation probability with the traditional procedure based on the Wightman function. The
remaining
sections are dedicated to the alternative calculation using the spherical modes. In Sec.~4 we obtain the
excitation probabilities for the final states of the field with a given number $\ell$ and discuss some of
their properties. In Sec.~5 we sum these probabilities and establish the connection with the result in Sec.~3.
Finally, in the last section we present the conclusions and make a few observations regarding the extension to
a similar calculation in a curved spacetime.

\section*{2. General formalism}

We recall here some general facts about the Unruh detectors. In its simplest form \cite{birr},  an Unruh
detector is a point-like system with a Hamiltonian $H_{det}$ that  is coupled to the quantum field via an
interaction Hamiltonian $H_{int} =\mu \varphi$, where $\mu$ is an operator responsible for the transitions
between different detector states, and $\varphi$ is the quantum field operator  at the position of the
detector. In the interaction picture, the Hamiltonian $H_{int}$ is
\bq
H_{int}(\tau) =\mu(\tau) \varphi[x(\tau)],
\eq
where $\tau$ is the detector's proper time, $\mu(\tau)$ is the transition operator in the interaction picture
\bq
\mu(\tau)=e^{iH_{det} \tau} \mu(0)\, e^{-iH_{det} \tau},
\eq
and $\varphi[x(\tau)]$ is the field operator evaluated  along the detector's trajectory $x(\tau)$.

Let us denote by $\varphi_K$ the quantum modes of the field. In the first order of the perturbation theory,
the transition amplitude\footnote{We will consider here only excitation amplitudes corresponding to energies
$E>0.$} from the ground state $E_0=0$ to an energy level $E$ of the detector and a final state $K$ of the
field is
\bq
{\cal A}_{0\rightarrow E,\, K}(\tau)=
\langle E\vert \mu(0) \vert E_0\rangle \times \int_{-\infty}^\tau d\tau^\prime e^{iE\tau^\prime}\varphi^*_K (x(\tau^\prime)),
\label{defa}
\eq
where we assumed that the interaction begins in the infinite past $\tau\rightarrow-\infty$. The corresponding
transition probabilities are
\bq
P_K(E,\tau)
= \vert {\cal A}_{0\rightarrow E,\, K}(\tau) \vert^2.
\eq

The total excitation probability if one ignores the final state of the field is given by the sum over modes
\bq
P(E,\tau)=\sum_{K} P_K(E,\tau).
\label{sumk}
\eq
The usual step at this point is to introduce the Wightman function
\bq
D^+(x, x^\prime)=\sum_{K}\varphi_K(x) \varphi_K^*(x^\prime).
\eq
in terms of which the sum  (\ref{sumk}) can be expressed as
\bq
P(E,\tau)=
\int_{-\infty}^\tau d\tau_1 \int_{-\infty}^\tau d\tau_2
\, e^{-iE(\tau_1-\tau_2)} D^+(x(\tau_1), x(\tau_2)).
\label{prow}
\eq

The vast majority of calculations of the detector's response are based on formula (\ref{prow}). However,
there are many cases in which one cannot obtain a closed form expression for $D^+(x, x^\prime)$, but one
can find the modes $\varphi_K(x)$. In these conditions, one can evaluate first the probabilities $P_K(E, \tau)$
and as a second step perform the sum over the modes. The aim of this paper is to show how such a calculations
works for an inertial trajectory in the Minkowski vacuum, when identifying $\varphi_K$ with the spherical
modes of the field.

\section*{3. Standard calculation}

We begin with the traditional method based on the Wightman function. As mentioned,
we use the noncovariant $i\varepsilon$ regularization, which consists in summing over the modes
with the convergence factor $e^{-\varepsilon \omega}$. The Wightman function in this case is
\bq
\qquad
D^+(x, x^\prime)=-\frac{1}{4\pi^2} \frac{1}{[(t-t^\prime-i\varepsilon)-({\bf x}-{\bf x}^\prime)]^2},
\quad \varepsilon >0.
\label{wf}
\eq
We take the trajectory of the detector to be (standard notation is used)
\bq
\qquad t(\tau)=\gamma \tau, \quad
{\bf x}(\tau)= \gamma \mbox{\boldmath $\beta$} \tau, \quad \beta \in (0,1).
\label{tra1}
\eq
Evaluating (\ref{wf}) along the trajectory (\ref{tra1}) one finds
\bq
D^+(\tau, \tau^\prime)=
-\frac{1}{4\pi^2} \frac{1}{(\Delta \tau-i \gamma \varepsilon)^2+\varepsilon^2(\gamma^2-1)}.
\label{nonwig}
\eq
Note that we keep in (\ref{nonwig}) the term $\sim \varepsilon^2$. As we will see, this term cannot be
ignored for a precise comparison with the result from the calculation using the spherical 
modes.

As a parenthesis,\, if one uses Schlicht's regularization \cite{schl}\, the Wightman function along the trajectory
is (\ref{nonwig}) with $\gamma=1$. This simply expresses the covariant nature of the regularization and the
Lorentz invariance of the Minkowski vacuum. We recall that in this procedure the regulator $\varepsilon$ is
essentially the length of the detector in its proper frame. We will also refer to this parameter
as the `length of the detector.'

The excitation probability  can be calculated as follows. Introducing (\ref{nonwig}) in the general
formula (\ref{prow}) one has
\bq
P_\varepsilon(E, \tau, \beta)=-
\frac{1}{4\pi^2}
\int_{-\infty}^\tau d\tau_1 \int_{-\infty}^\tau d\tau_2
\frac{e^{-i E (\tau_1-\tau_2)}}{((\tau_1-\tau_2)- i \gamma \varepsilon)^2}.
\label{i1}
\label{casint}
\eq
Following Ref.~\cite{svai} we introduce the new variables
\bq
\eta =2\tau-\tau_1-\tau_2,
\nonumber
\\
\xi=\tau_1-\tau_2,\,\,\,
\quad\quad
\label{cvar}
\eq
in terms of which (see Fig.~1)
\bq
P_\varepsilon(E, \beta)=-
\frac{1}{8\pi^2}
\int^{\infty}_0 d\eta \int_{-\eta}^{+\eta}
d\xi \frac{e^{-i E \xi}}
{(\xi-i \gamma \varepsilon)^2+\varepsilon^2(\gamma^2-1)}.
\label{petaxi}
\eq
One sees that the dependence on $\tau$ has disappeared from the integral, which is a consequence of the
time translational invariance of the system. In the presence of a decoupling of the interaction, this
would not be the case. However, as we point out below, in the adiabatic limit the decoupling factor
will make no difference.

\begin{figure}

\centerline{\includegraphics[width=5.5in, height=4in, angle=0]{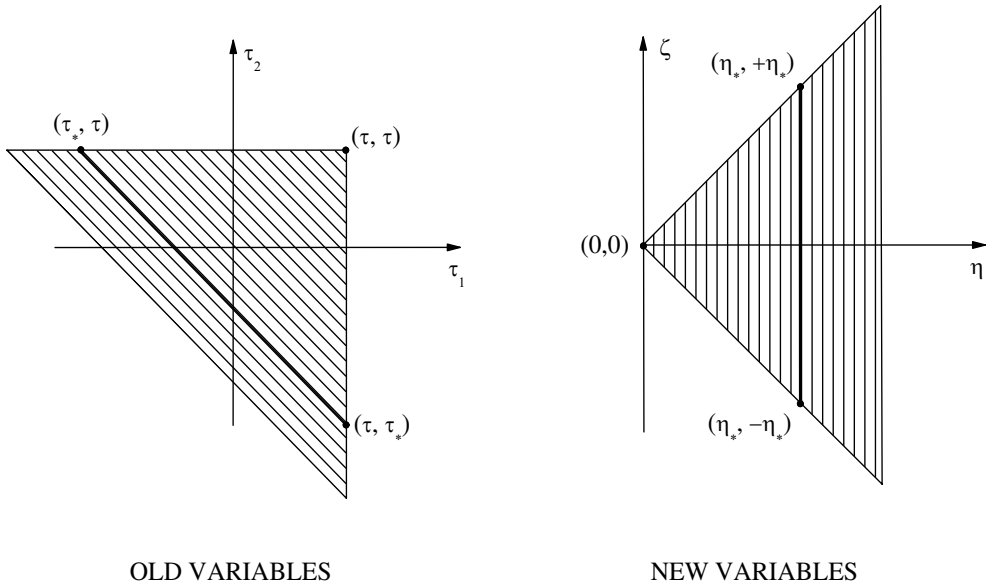}}

\caption{The integration domain in (\ref{casint}) in terms of the variables $(\eta, \xi)$.}

\end{figure}

The integral (\ref{petaxi}) can be simplified with the following trick. Taking
advantage of the fact that the integrand
does not depend on $\eta$, we can write (the prime means derivation with respect to $\eta$):
\bq
P_\varepsilon(E, \beta)&=&\lim_{{\bf \eta} \rightarrow \infty}\int^{\bar\eta}_0 d\eta
\left\{\eta^\prime \int_{-\eta}^{+\eta}
d\xi f(\xi)\right\}
\nonumber
\\
&=&\lim_{\bar\eta\rightarrow \infty}\left\{  \bar \eta \int_{-\bar \eta}^{+\bar\eta}
d\xi\, f(\xi)
-
\int^{\bar\eta}_0 d\eta\, \eta \,[f(\eta) -f(-\eta)]
\right\}
\nonumber
\\
&=&\lim_{\bar\eta\rightarrow \infty} \left\{\int_{-\bar \eta}^{+\bar\eta} d\xi\,
(\bar \eta -\vert \xi \vert)
f(\xi)\right\}.
\label{using}
\eq
The second identity follows from an integration by parts, and the third identity with a
rearrangement of the integral with respect to $\eta$. Applying the procedure in (\ref{petaxi}) we find
\bq
P_\varepsilon(E, \beta)=
-\frac{1}{8\pi^2}
\lim_{\bar\eta\rightarrow \infty} \int_{-\bar \eta}^{+\bar\eta} d\xi\,
\frac{(\bar \eta -\vert \xi \vert)}
{(\xi-i \gamma \varepsilon)^2+\varepsilon^2(\gamma^2-1)}\,e^{-i E \xi}.
\label{intep}
\eq
A further simplification is obtained by integrating with respect to
\bq
\xi \rightarrow \xi/E,
\eq
which finally yields
\bq
P(\varepsilon E, \beta)=
-
\frac{1}{8\pi^2}
\lim_{\bar\eta\rightarrow \infty} \int_{-\bar \eta}^{+\bar\eta} d\xi\,
\frac{(\bar \eta  -\vert \xi \vert)}
{(\xi-i \gamma \varepsilon E)^2+(\varepsilon E)^2(\gamma^2-1)}\,e^{-i \xi}.
\label{inteq}
\eq
Notice that the result depends only on the combination $\varepsilon E$, as could have been guessed from
dimensional considerations.

An analytical expression for (\ref{inteq}) seems hard to obtain. The important fact is that the limit
exists\, and it is finite, which can be easily checked with a numerical calculation. A particularly
simple form for (\ref{inteq}) can be found \cite{svai} for a sufficiently small length of the detector
$\varepsilon E\ll 1$. The result is (see the Appendix):
\bq
P(\varepsilon E, \beta)\simeq \frac{1}{4\pi^2} \ln (\gamma \varepsilon E)^{-1} +c,
\label{prolim}
\eq
where $c$ is a constant of the order of unity.

As expected, the probability (\ref{prolim}) diverges for $\varepsilon\rightarrow 0$. Formally,
this is a consequence of the divergent behavior of the Wightman function in the coincidence limit. From a more
physical point of view, the divergence reflects the unphysical  idealization implied by a pointlike detector
and the sudden decoupling of the interaction \cite{pere, ramk}. A representation for (\ref{inteq}) will be
given in Sec.~6, when we make contact with the calculation based on the spherical modes.

Note that we have not used a  decoupling of the interaction: introducing a decoupling factor and taking at
the end of the calculation the adiabatic limit would lead to the same result. This should be immediate from the
fact that the integral (\ref{inteq})  is already finite. If one wants, the role of the adiabatic decoupling in
the far past in our case is played by the fact that the Wightman function vanishes sufficiently fast when one of
the points is at $\tau\rightarrow -\infty$, so that no extra decoupling is necessary.

We have to stress that our result  (\ref{inteq}) is smaller by a factor of 2 than the excitation probability for
a sudden switch on of the interaction in the infinite past \cite{svai}. More exactly, the excitation probability
\cite{svai} for the inertial detector suddenly coupled to the field at $\tau_0=0$ and abruptly decoupled
at $\tau>0$ is given by two times\footnote{In\, Ref.~\cite{svai} the term $\sim\varepsilon^2$ in
the Wightman function is ignored; the above statement is exact if one keeps this term. Anyway, the particular
form of $D^+$ is irrelevant; it is the form of the integrals in the $\eta$-$\xi$ space that matters.} the integral
(\ref{inteq}) with the identification
\bq
\bar\eta\equiv\tau.
\eq
The factor $1/2$ is directly related to the adiabatic coupling in the infinite past.
In brief, the explanation for this factor is as follows: for a sudden switch-on of the interaction
at $\tau_0>-\infty$, the domain of integration in (\ref{casint}) is the square $[\tau_0, \tau]\times [\tau_0, \tau]$,
where as a consequence of the $\eta$-independence of the integrand the  triangles above and under the second diagonal give
the same contribution (see Fig.~4 in Ref.~\cite{svai}). If one assumes that the interaction is adiabatically
decoupled at $\tau_0 \rightarrow -\infty$, the contribution of the lower triangle corresponding to large times in the past
is eliminated due to the decoupling factor. This leaves us with the semi-infinite triangle (i.e. half of the square)
in Fig.~1, and hence the factor $1/2$ in our result.

\section*{4. The multipole probabilities}

We now consider the calculation based on  the spherical modes. The first step is to determine the  transition
amplitudes (\ref{defa}). The spherical modes for the scalar field are (with the conventional normalization for
the Klein-Gordon field)
\bq
\varphi_{\omega \ell m}(t, r, \theta, \phi)=\sqrt{\frac{\omega}{\pi}} j_\ell\,(\omega r)
Y_{\ell\,m}(\theta, \phi),
\eq
where $j_\ell$ are the spherical Bessel modes of the first kind and $Y_{\ell m}$  are the spherical harmonics. An
obvious way to simplify things is to choose the detector's trajectory along the $z$-axis, i.e.
\bq
x=0,\quad y=0,\quad z(\tau)=\gamma \beta \tau,
\eq
in which  case only the amplitudes with $m=0$ survive. We recall that at a space inversion the spherical
harmonics pick up a phase $\eta=(-1)^\ell$, and that the same factor relates the functions $j_\ell$ when
changing the sign of the argument. It is possible then to formally define the trajectory as
\bq
\qquad
r(\tau)=\gamma \beta \tau,
\quad \tau \in (-\infty, \infty),
\eq
and use for the spherical harmonics the expression for $\theta=0$, i.e.
\bq
Y_{\ell\,0}(0, \phi)= \sqrt{\frac{2\ell+1}{4\pi}}.
\eq
Introducing the relations above in (\ref{defa}), one finds that the amplitudes are
\bq
{\cal A}_{0\rightarrow E,\, \omega\ell}\,(\tau)= \frac{1}{2\pi }\sqrt{(2\ell +1)\omega}
\int_{-\infty}^{\tau} d\tau^\prime e^{i(E+\gamma \omega)\tau^\prime}
j_\ell (\gamma\omega\beta \tau^\prime),
\label{ampmin1}
\eq
where we neglected as usual the matrix element of the operator $\mu(0)$.

From now we will start to rely on numerical calculations. A major problem when numerically evaluating
(\ref{ampmin1})  is the highly oscillatory behaviour of the integrand for $\tau^\prime \rightarrow -\infty$
at large frequencies $\omega$. This  can be remedied with an integration in the complex plane. It is not hard
to see that the integrand in (\ref{ampmin1}) is an analytic function of $\tau^\prime$ that vanishes\footnote{Conditions
$E>0$ and $\beta<1$ are essential  at this point.} for Im $\tau^\prime \rightarrow -\infty$, so that
the integration contour can be rotated to the semi-infinite line
\bq
\qquad \tau^\prime=\tau-is, \quad s\in [0, \infty).
\eq
It is further convenient to rescale the integration variable as $s\rightarrow s/(\gamma \omega)$, which
brings (\ref{ampmin1}) to the form
\bq
{\cal  A}_{0\rightarrow E,\, \omega\ell}\,(\tau)=
\frac{i}{2\pi \gamma}
\sqrt{\frac{2\ell+1}{\omega}}
\int_{0}^{\infty} ds\,
e^{-(s+i \gamma \omega \tau)\left(\frac{E}{\gamma\omega}+1\right)} j_\ell (\gamma\omega \beta \tau-i\beta s).
\label{ampz}
\eq
The integral is now rapidly convergent due to the factor $e^{-s}$. Formula (\ref{ampz}) is the
basis for the numerical results presented below.

The excitation probabilities corresponding to  (\ref{ampz}) are
\bq
P_\ell(E, \tau; \omega)=\vert{\cal A}_{0\rightarrow E,\, \omega\ell}\,(\tau) \vert^2.
\eq
Ignoring the frequencies in the final states of the field, the excitation probabilities
are
\bq
P_\ell(E, \tau)=\int_0^\infty d\omega P_\ell(E, \tau; \omega).
\label{pa}
\eq
Explicitly, the integrals (\ref{pa}) are given by (notice the dependence on $E\tau$):
\bq
P_\ell(E, \tau)\qquad\qquad\qquad\qquad\qquad\qquad\qquad\qquad\qquad\qquad\qquad\qquad\qquad
\nonumber
\\
\qquad\qquad
=\frac{2\ell+1}{4\pi^2 \gamma^2}\int_0^\infty \frac{d\bar \omega}{\bar\omega}
\Big\vert
\int_{0}^{\infty} ds\,
e^{-(s+i \gamma \bar \omega E\tau)\left(\frac{1}{\gamma\bar\omega}+1\right)} j_\ell (\gamma\bar\omega \beta E\tau-i\beta s)
\Big\vert^2,
\eq
where we integrated with respect to the dimensionless frequency
\bq
\bar \omega =\omega/E.
\eq
We will call (\ref{pa}) the `multipole probabilities.' We now take a closer look at these probabilities.

To begin with, it is useful to remark that the amplitudes (\ref{ampz}) can be given a direct physical significance
in the following way. Let us imagine a large number $N$ of identical detectors uniformly distributed on a sphere,
having its\, radius equal to the radial coordinate of the detector along the trajectory, $r(\tau)=\gamma \beta \vert
\tau\vert$. Let us further consider that these detectors somehow interact between themselves in such a way that
they form a `global detector,' whose transition amplitude is given by the coherent sum of the individual detectors.
It is then easy to see that for $N\rightarrow \infty$ the transition amplitude of the global detector
is proportional to the amplitude $\ell=0$ defined by (\ref{ampz}). A similar interpretation can be given to the
amplitudes with an arbitrary $\ell$, choosing on the sphere a density of detectors  $\sim Y_{\ell 0} (\theta, \phi)$.

A representation of the probabilities $P_\ell(E, \tau)$ as function of $E\tau$ for
$\ell=0,1$ and 2 and different velocities of the detector is
shown in Figs.~2-4. Some comments are in order. First, the probabilities depend on time. This might be
surprising, given the inertial trajectory of the detector and the vacuum state of the field. However, the property
can be naturally understood if one has in mind the time-dependent geometry of the spherical detectors introduced above.

\begin{figure}

\centerline{\includegraphics[width=5.3in, height=3.5in, angle=0]{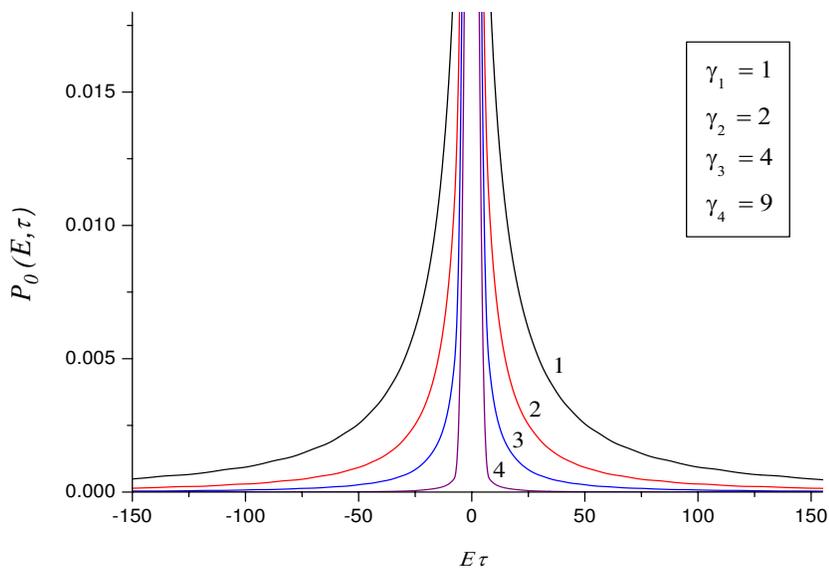}}

\caption{The excitation probability $P_{\ell=0}(E, \tau)$ as a function of $E\tau$ for different values of
the $\gamma$ factor. The curves are symmetric under the time reflection $\tau \rightarrow -\tau$;
see (\ref{haro}).}

\end{figure}

\begin{figure}

\centerline{\includegraphics[width=5.3in, height=3.5in, angle=0]{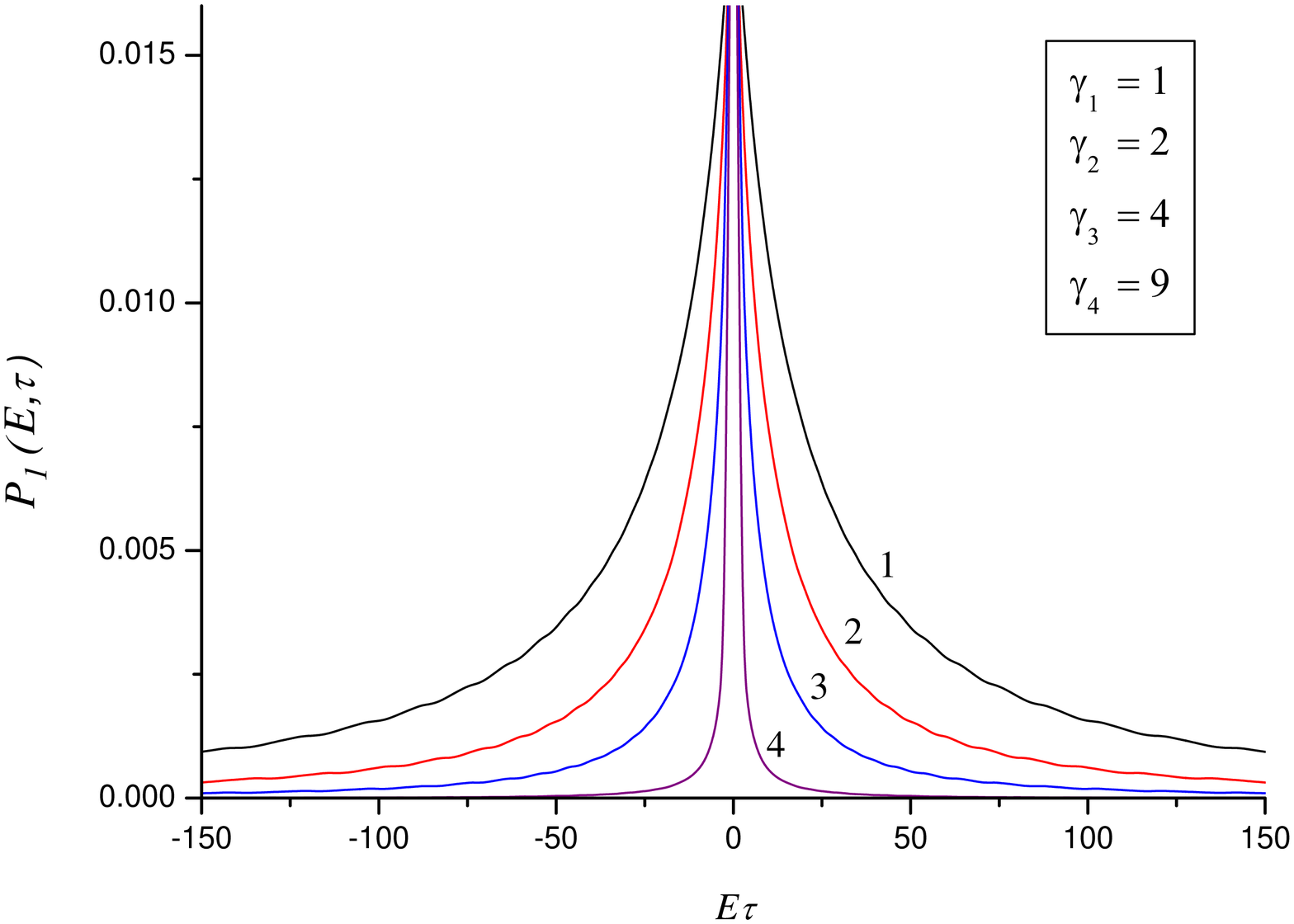}}

\caption{The same as in Fig.~2 for the probability $P_{\,\ell=1}(E, \tau)$.}

\end{figure}

\begin{figure}

\centerline{\includegraphics[width=5.3in, height=3.5in, angle=0]{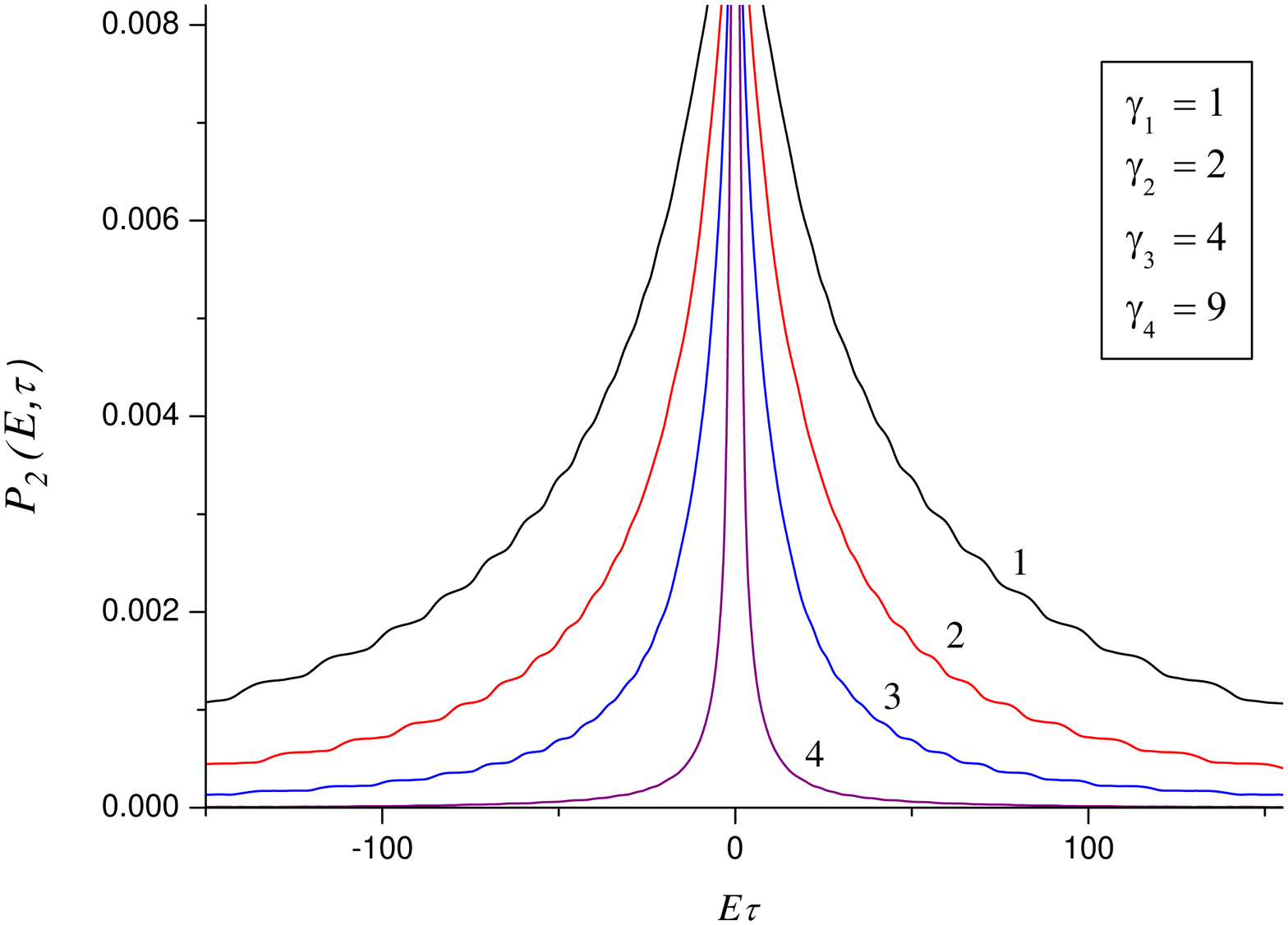}}

\caption{The same as in Fig.~2 for the probability $P_{\ell=2}(E, \tau)$.}

\end{figure}

Another relevant fact is that the probabilities vanish at infinite times $\tau\rightarrow \infty$. This can be
proven observing that\footnote{Relation (\ref{haro}) easily follows from taking the complex
conjugate of (\ref{ampmin1}).}
\bq
P_\ell(E, \tau)=P_\ell(E, -\tau),
\label{haro}
\eq
which makes the property immediate since the amplitudes vanish in the infinite past
$\tau\rightarrow -\infty$. The vanishing of the excitation probabilities at $\tau\rightarrow \infty$ is clearly a
reminiscence of the fact that the  (pointlike) detector does not get excited along an inertial trajectory.

Note also the large probabilities near the time $\tau=0$. In fact, the probabilities diverge at this point.
This is a consequence of the fact that for $\tau=0$ the dependence on $\omega\rightarrow\infty$ disappears
from the integrand in (\ref{ampz}). This makes the amplitudes at large frequencies behave as $\sim 1/\sqrt{\omega}$,
in which conditions the integral over frequencies  (\ref{pa}) diverges. Physically, the divergence can be associated
to the singular geometry of the detector sphere at $\tau=0$, in which case all $N$ detectors collapse into a point.

Rather counterintuitively, one sees that the probabilities decrease with the velocity $\beta$. Formally, this
is the effect of the inverse relativistic factor $1/\gamma$ in front of the integral (\ref{ampz}). Physically, the
decreasing behavior with $\beta$ could be understood as follows. First, one can naturally admit that the main
contribution in the excitation of the spherical detector comes from the interval $\Delta\tau_{exc}$ which
implies a significant variation of the geometry of the sphere. The geometry of a spherical surface can be
described by the curvature
$\kappa=1/r$, so the significant `variation rate of the geometry' would be $d\kappa/d\tau \sim \beta/r^2$. One can
conclude from here that the excitation is mainly due to the interval in which the radius of the sphere takes
very small values, i.e. $r_{exc} \simeq 0$. A large velocity implies a small such interval, and hence
the desired `explanation.'

\section*{5. Summing over the multipoles}

An  immediate question when considering the curves in Fig.~2 is: how the time-dependent probabilities
$P_\ell(E, \tau)$ are compatible with the time-independent excitation probability for an inertial
trajectory of the detector? We now  clarify this point. It is not hard to guess that the answer lies in
the sum over $\ell$.

The idea is to look at the picture in the frequency space. Let us introduce the partial multipole sums
\bq
{\cal P}_L(E, \tau; \omega)=\sum_{\ell=0}^
LP_\ell(E, \tau; \omega).
\label{sump}
\eq
It turns out that it is convenient to consider the dependence with respect to
\bq
\sigma \equiv \ln \omega,
\eq
and redefine  (\ref{sump}) as
\bq
{\cal P}_ L(E, \tau; \sigma)\equiv\, e^\sigma\,  {\cal P}_ L(E, \tau; e^\sigma) .
\label{sums}
\eq
In these
conditions (\ref{sump}) integrated over the frequencies is
\bq
{\cal P}_L(E, \tau)\equiv \int_0^\infty d\omega {\cal P}_L(E, \tau; \omega)
\nonumber
\\
=\int_{-\infty}^\infty d\sigma {\cal P}_L(E, \tau; \sigma)
\nonumber
\\
=\sum_{\ell=0}^L P_\ell(E, \tau).\qquad\,\,\,\,
\label{ir}
\eq
Explicitly, the quantity under the  $\sigma$-integral is
\bq
{\cal P}_ L(E, \tau, \sigma)
=\frac{1}{4\pi^2 \gamma^2}\sum_{\ell =0}^L
 (2\ell+1)
\qquad\qquad\qquad\qquad\qquad\qquad\quad
\nonumber
\\
\times {\Big \vert}
\int_{0}^{\infty} ds\,
e^{-\left(s+i \gamma\tau e^{\sigma} \right) \left(\frac{E}{\gamma}e^{-\sigma}+1\right)}
j_\ell (\gamma\beta \tau e^{\sigma}-i\beta s) {\Big \vert}^2.
\label{pes}
\eq

We now consider more closely (\ref{pes}). Figure 5 shows the typical form of  ${\cal P}_ L(E, \tau, \sigma)$ as a
function of $\sigma$ for different values of $L$ at a fixed $E$ and $\tau$. Note that the areas below the curves
are the probabilities ${\cal P}_L(E, \tau)$. One sees that each $L$ comes with a cutoff $\sigma_{cut} (L)$,
which increases when $L$ increases. The key fact is that in the limit $L\rightarrow \infty$\, the cutoff disappears,
and for high enough frequencies the curves become a horizontal line,
\bq
\qquad
\lim_{L\rightarrow \infty}{\cal P}_ L(E, \tau, \sigma)\simeq \mbox {constant},
\qquad \sigma \gg 1.
\eq

\begin{figure}

\centerline{\includegraphics[width=5.3in, height=3.7in, angle=0]{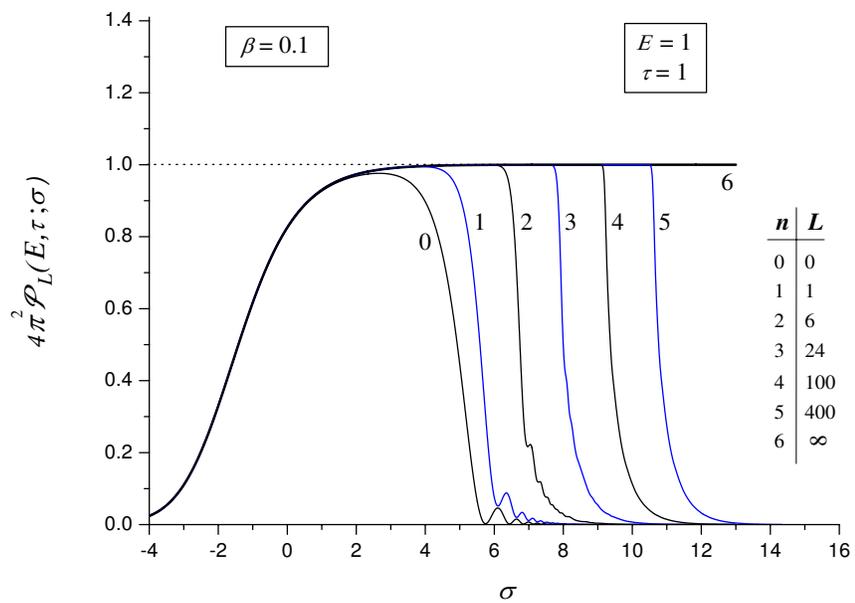}}

\caption{The probability ${\cal P}_ {L}\,(E, \tau, \sigma)$ in units of $4\pi^2$  shown as a function
of $\sigma$ for different numbers $L$ and fixed values of $E$, $\tau$ and $\beta$. The cutoff in the limit
$L\rightarrow \infty$ is pushed at $\sigma\rightarrow \infty$.}

\end{figure}

The nice thing is that for $L\rightarrow \infty$ the picture is the same for all $\tau$. The mechanism how the sum
does this is shown in Figs.~6 and 7. It can be observed that the dependence on $\tau$ is practically encoded in
the cutoff
$\sigma_{cut} (L, \tau)$. Increasing the value of $\vert\tau\vert$ with $L$ fixed decreases $\sigma_{cut} (L, \tau)$,
but it leaves the curves at lower frequencies unchanged. Essentially, an increasing $L$ has the opposite effect
on $\sigma_{cut} (L, \tau).$ The limit $L\rightarrow \infty$ pushes the cutoff at infinity, eliminating thus
completely the dependence on $\tau$. This basically answers our
question.

\begin{figure}

\centerline{\includegraphics[width=5.3in, height=3.7in, angle=0]{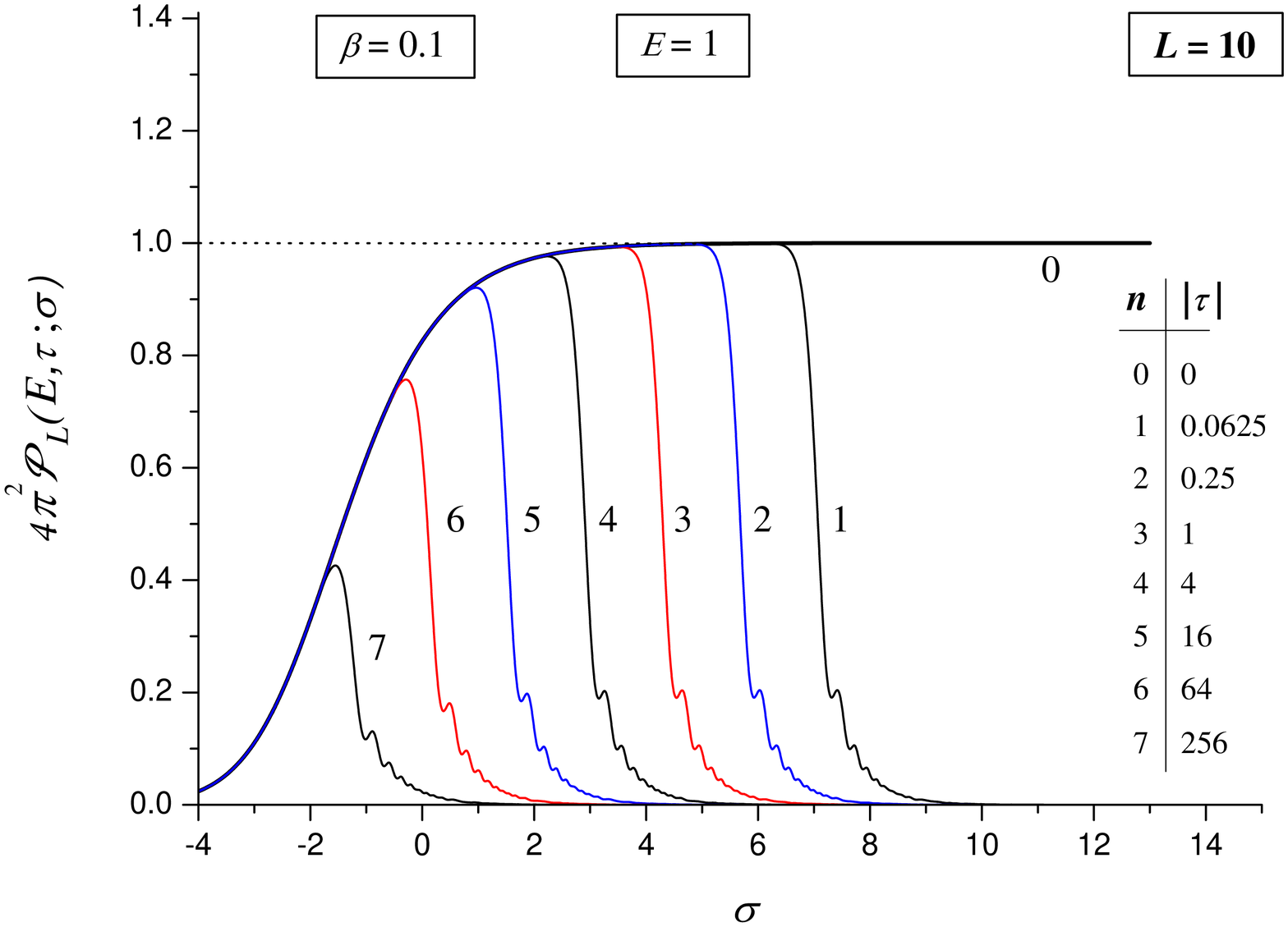}}

\caption{The probability ${\cal P}_ {L=10}\,(E, \tau, \sigma)$ in units of $4\pi^2$  shown as a function
of $\sigma$ for different values of $\vert \tau\vert$ and $E$, $\beta$ fixed. The solid curve $n=0$
coincides with the limit curve $L\rightarrow \infty$. This curve is independent of $\tau$.}

\end{figure}

\begin{figure}

\centerline{\includegraphics[width=5.5in, height=4in, angle=0]{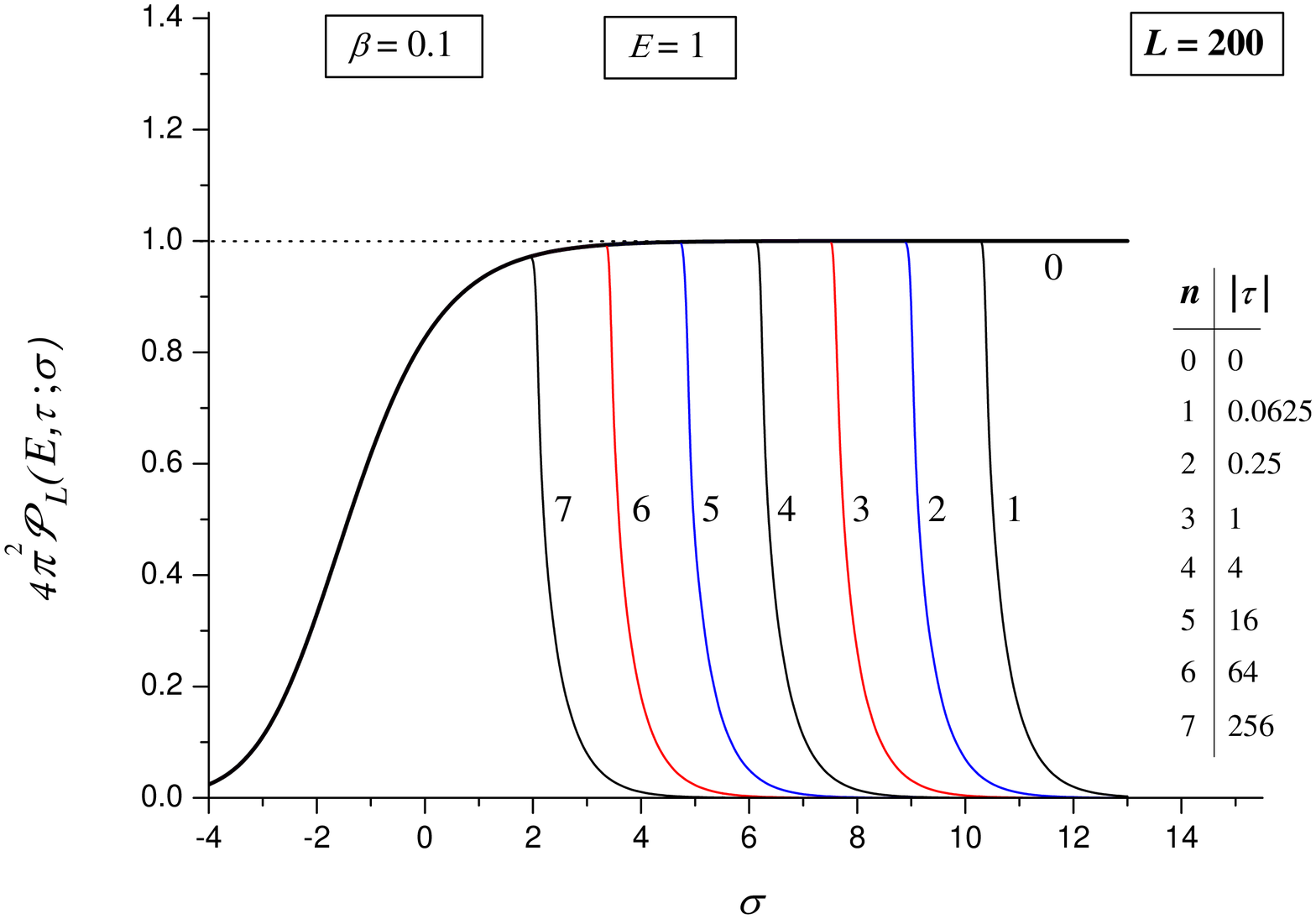}}

\caption{The same as in Fig.~6 for ${\cal P}_ {L=200}\,(E, \tau, \sigma)$. Notice the displacement
of the cutoffs towards larger values of $\sigma$ for the curves $n=1..7$.}

\end{figure}

Note from the plots that for the curves $L\rightarrow \infty$  the limit value at large frequencies
is
\bq
\lim_{L\rightarrow \infty} 4\pi^2{\cal P}_ L(E, \tau, \sigma\rightarrow \infty)=1.
\label{duh}
\eq
We will prove this  relation for  $\tau=0$ in the next section. For arbitrary times, we trust the
numerical calculations.

\section*{6. The full excitation probability}

We arrive now to the task of recovering the excitation probabilities obtained in Sec.~3.
We have to perform the integral over frequencies (\ref{ir}) with $L\rightarrow \infty$. It is clear
from the plots above that the integrals diverge due to the contributions from $\sigma \rightarrow\infty$,
so that a cutoff is needed. In order to make contact with the previous result, we obviously have to
use the same convergence factor $e^{-\varepsilon\omega}$. We have thus to evaluate
\bq
{\cal P}_\varepsilon(E, \beta)\equiv
\lim_{L\rightarrow \infty}\int_0^\infty d\omega{\cal P}_L(E, \tau; \omega)\,  e^{-\varepsilon \omega}
\,\,\,\,
\nonumber
\\
=
\lim_{L\rightarrow \infty}\int_{-\infty}^\infty d\sigma {\cal P}_L(E, \tau; \sigma )\,  e^{-\varepsilon e^\sigma}.
\label{carta}
\eq

We first determine ${\cal P}_ L(E, \tau, \sigma)$ in the limit $L\rightarrow \infty$. We trust the independence of
$\tau$ evidenced by the numerical calculation, so that it is sufficient to pick up a special time $\tau_*$. Choosing
$\tau_*=0$, the sum simplifies to (see (\ref{pes}))
\bq
\lim_{L\rightarrow \infty}
{\cal P}_ L(E, \tau, \sigma)
\qquad\qquad\qquad\qquad\quad\qquad\quad\qquad\qquad\quad\quad
\nonumber
\\
=\frac{1}{4\pi^2\gamma^2}\sum_{\ell =0}^\infty
 (2\ell+1){\Big \vert}
\int_{0}^{\infty} ds\,
e^{-s \left(\frac{E}{\gamma}\,e^{-\sigma}+1\right)} j_\ell (-i\beta s) {\Big \vert}^2.
\label{pes1}
\eq
A further simplification occurs if one considers the high-frequency limit $\sigma \rightarrow \infty$, in which case
\bq
\lim_{L\rightarrow \infty}
{\cal P}_ L(E, \tau, \sigma\rightarrow \infty)
\qquad\qquad\qquad\quad\qquad\quad\qquad\quad\quad\quad
\nonumber
\\
=\frac{1}{4\pi^2 \gamma^2}\sum_{\ell =0}^\infty
 (2\ell+1) {\Big \vert}
\int_{0}^{\infty} ds\,
e^{-s} j_\ell (-i\beta s) {\Big \vert}^2.\quad\qquad\quad
\label{pes2}
\eq
The magic formula which solves (\ref{pes2}) is:
\bq
\sum_{\ell =0}^\infty
 (2\ell+1){\Big \vert}
\int_{0}^{\infty} ds\,
e^{-s} j_\ell (-i\beta s) {\Big \vert}^2=\gamma^2.
\label{pes3}
\eq
This can be easily checked with a brute force method by using the power expansion of the Bessel functions and
reorganizing the result in powers of $\beta$. The identity (\ref{duh}) is now immediate
from (\ref{pes3}).

The useful relation for summing (\ref{pes1}) is obtained by making in (\ref{pes3}) the rescalings
$s \rightarrow \alpha s$, $\beta\rightarrow \beta/\alpha$, from which ($\alpha >\beta$)
\bq
\sum_{\ell =0}^\infty
 (2\ell+1){\Big \vert}
\int_{0}^{\infty} ds\,
e^{- \alpha s} j_\ell (-i\beta s) {\Big \vert}^2=\frac{1}{\alpha^2-\beta^2}.
\eq
Applying to (\ref{pes1}) one finds
\bq
\lim_{L\rightarrow \infty}
{\cal P}_ L(E, \tau, \sigma)=\frac{1}{4\pi^2}\frac{1}{\left( \frac{E}{\gamma}\,e^{-\sigma}+\gamma\right)^2-\gamma^2\beta^2}.
\label{uhuru}
\eq
Finally, introducing (\ref{uhuru}) in the second integral in (\ref{carta}) and integrating with respect to
$\sigma\rightarrow \sigma +\ln (E/\gamma)$ the total probability is (notice the \,dependence on $\varepsilon E$)
\bq
{\cal P}( \varepsilon E, \beta)=\frac{1}{4\pi^2} \int_{-\infty}^\infty d\sigma
\frac{e^{-(\varepsilon E/\gamma) e^\sigma}}{\left(e^{-\sigma}+\gamma\right)^2-\gamma^2\beta^2}.
\label{ints}
\eq

A representation of the integrand in (\ref{ints}) as a function of $\sigma$ for different values
of $\varepsilon E$ is shown in Fig.~8. From the double exponential in the long fraction, it is clear
that we have a cutoff at
\bq
\sigma_{CUT} \simeq
\ln (\varepsilon E/\gamma)^{-1}.
\label{cuts}
\eq
Combining this with the unit limit of the denominator for $\sigma \rightarrow \infty$, one can read that
for a sufficiently small $\varepsilon E$ the integral diverges as
\bq
{\cal P}(\varepsilon E, \beta)\simeq \frac{1}{4\pi^2} \ln (\varepsilon E/\gamma)^{-1}.
\label{raw}
\eq
This is identical with the dependence on $\varepsilon E$ in (\ref{prolim}). The discrepancy in the
terms $\sim \ln \gamma$ has to be ascribed to the raw evaluation in (\ref{raw}).

\begin{figure}

\centerline{\includegraphics[width=5.3in, height=3.5in, angle=0]{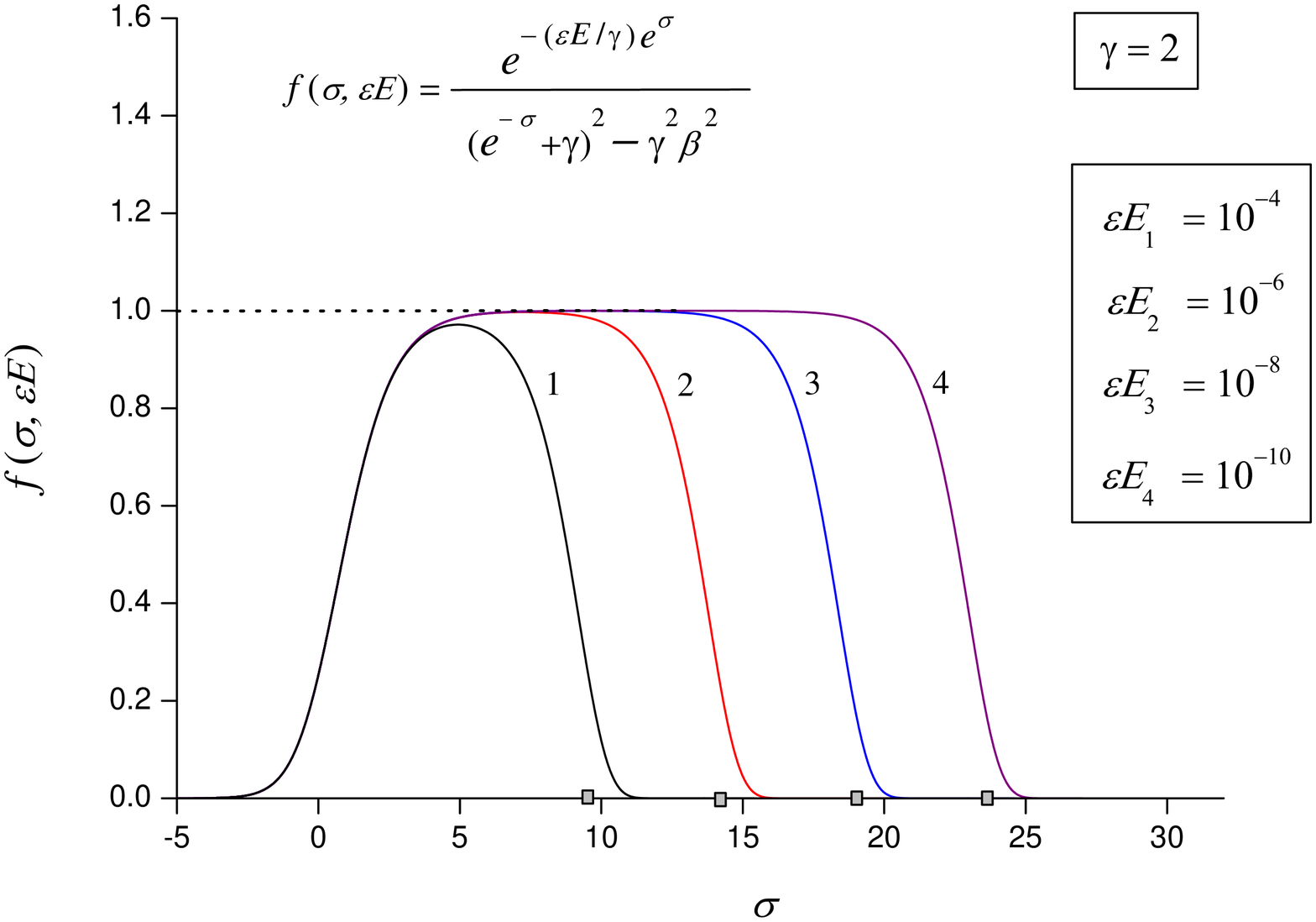}}

\caption{The integrand in (\ref{ints}) shown as a function of $\sigma$ for different values of $\varepsilon E$
and $\gamma=2$. The squares on the $x$-axis indicate the cutoffs defined by (\ref{cuts}).}

\end{figure}

We do not have an analytical expression for (\ref{ints}). The crucial fact that can be inferred from numerical
calculations is that (\ref{ints}) coincides indeed\footnote{Using \texttt{NIntegrate} in the $Mathematica$ software
the difference between the two integrals for the values in Fig.~9 is at most $\sim 10^{-4}$. Increasing the precision
in \texttt{ NIntegrate} the difference can be decreased by many orders of magnitude.} with the previous result
(\ref{inteq}). A representation of ${\cal P}(\varepsilon E, \beta)$ as a function of $\varepsilon E$ for different
velocities is shown in Fig.~9. This concludes our calculation.

\begin{figure}

\centerline{\includegraphics[width=5.2in, height=3.5in, angle=0]{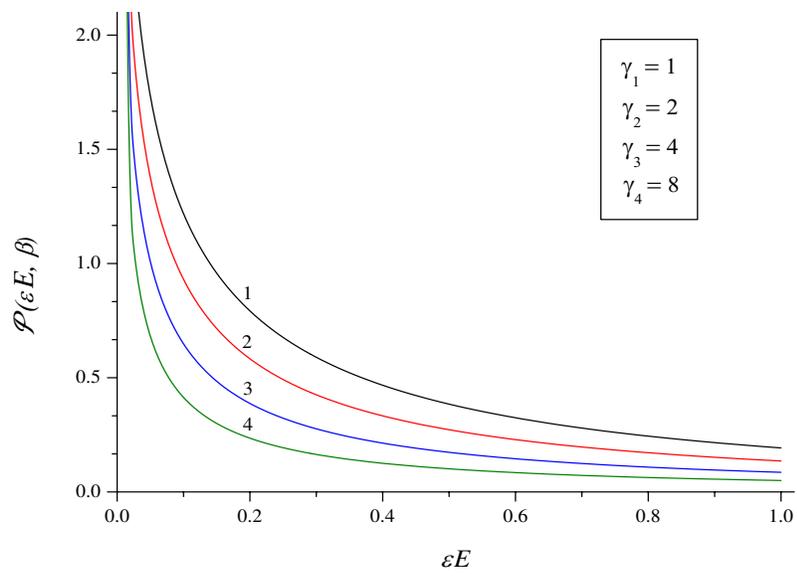}}

\caption{The
excitation probability of the detector as a function of $\varepsilon E$ for different values of the factor
$\gamma$. The curves defined by the integrals (\ref{inteq}) and (\ref{ints}) coincide.}

\end{figure}

For curiosity, in Fig.~10\, we represented the probabilities defined by the integral (\ref{inteq}) with
the term $\sim (E\varepsilon)^2$ in the Wightman function ignored. Notably, one sees that for values as
small as $\varepsilon E \sim 10^{-4}$ the differences with respect to the exact result can be significative.

\begin{figure}

\centerline{\includegraphics[width=5.2in, height=3.5in, angle=0]{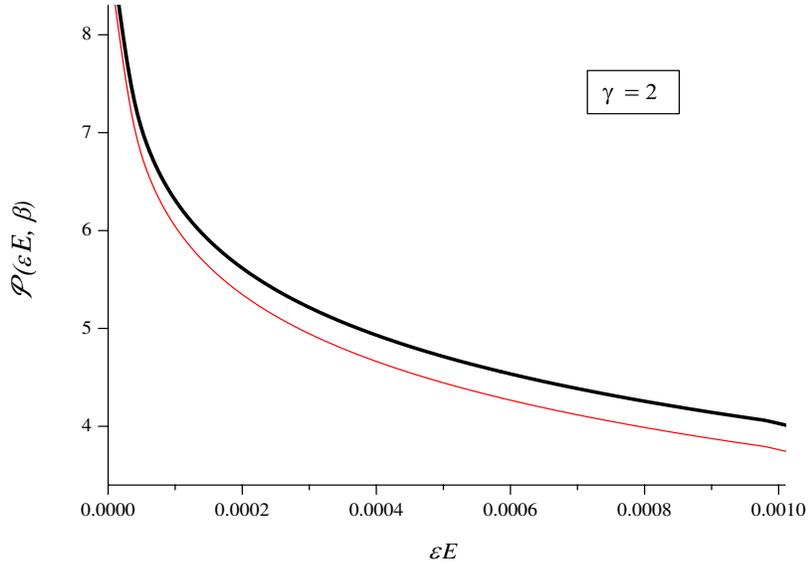}}

\caption{The exact excitation probability (solid line)  compared with the approximate
value defined by expression  (\ref{inteq}) with the
term $\sim  \varepsilon^2$ in the Wightman function ignored (thin line). The gamma factor is $\gamma=2$.}

\end{figure}

\section*{7. Conclusions}

The traditional way to calculate the response of the Unruh detector is by using an integral over the
Wightman function of the field. This implies summing first over the modes, and then integrating
along the trajectory of the detector. We considered here the response of an inertial detector in the
Minkowski vacuum, and presented a calculation via a different route: we  obtained first the excitation
probabilities for the final states of the field with one particle per mode, using the particle states defined
by the spherical modes, and then we summed over the modes. Integrating with a frequency cutoff $e^{-\varepsilon \omega}$
we recovered, as expected, the result based on the Wightman function with the $i\varepsilon$ regularization.
Explicitly, the integrals that define the excitation probabilities in the two calculations are:
\bq
\qquad\qquad
-
\frac{1}{8\pi^2}
\lim_{\bar\eta\rightarrow \infty} \int_{-\bar \eta}^{+\bar\eta} d\xi\,
\frac{(\bar \eta  -\vert \xi \vert)}
{(\xi-i \gamma \varepsilon E)^2+(\varepsilon E)^2(\gamma^2-1)}\,e^{-i \xi}
\nonumber
\\
\nonumber
\\
\qquad
=
\frac{1}{4\pi^2} \int_{-\infty}^\infty d\sigma
\frac{e^{-(\varepsilon E/\gamma) e^\sigma}}{\left(e^{-\sigma}+\gamma\right)^2-\gamma^2\beta^2}.
\qquad\qquad
\quad
\eq

We carried out this exercise mainly as a preliminary step for a similar calculation of the response of the
detector in a curved spherically symmetric space. Assuming, as here, a radial inertial trajectory,  the
extension to a curved background should be unproblematic. The only essential difference will consist in the form of
the radial functions in the quantum modes. It is reassuring to know that in Minkowski space we recover the
correct result.

A few points to be emphasized
are as follows. The excitation probability obtained here correspond
to a sudden switch off of the interaction and an adiabatic switch on in the infinite past. Although we have not
introduced a decoupling factor, the last property is implicit in our calculation since the interaction begins
at $\tau\rightarrow -\infty$, and because in this limit the integrands in the $\tau$-integrals
vanish (this is valid in both procedures). Recognizing  this fact is essential for a correct physical interpretation
of the result. A similar calculation for a sudden switch on of the interaction in the infinite past \cite{svai}
leads to a probability larger by a factor of two.

An interesting finding was that, despite the inertial trajectory of the detector, the excitation probabilities
for the final field states of the field with a given number $\ell$ are time-dependent quantities. However,
after performing the multipole sum, the time dependence disappears. In brief, this is ensured by the following
mechanism: Each multipole sum over $\ell \leq L$ comes in the log-frequency space $\sigma=\ln\omega$\,
with a well-defined cutoff $\sigma_{cut}(L, \tau)$, which practically encodes the dependence on time.
The limit $L\rightarrow \infty$ sends the cutoff to infinity, eliminating the dependence on $\tau$.

We end with a couple of observations, having in mind the extension of our procedure to the calculation
of the detector's response in a spherically symmetric curved spacetime. A lesson to be learned from
our analysis is as follows. Due to
general considerations, one expects the picture in the $\sigma$-space at large frequencies to be the
same as that in the Minkowski space. This means that it will possible to naturally separate the excitation
probability into (1) a Minkowski component, which will contain the contribution of high frequencies and
the dependence on the regulator $\varepsilon$, and (2) a low frequency contribution, which will be
independent of $\varepsilon$ and which will encode the effect of the curved background.

The key fact is that, as strongly indicated by the plots in Sec.~5, in order to obtain the low frequencies
contribution it will be sufficient to sum, without loss of precision, up to a not very large value of $L$.
This is a desirable simplification, since numerical calculations involving modes with large  numbers $\ell$
can be  problematic.\footnote{For example, a problem could be as follows. In the asymptotic region of a spherically
symmetric space, the radial functions are generally given by a combination of the spherical Hankel functions
$\sim H_{\nu}^{(1, 2)}(\omega  r)$ with $\nu=\ell+\frac{1}{2}$. Suppose that we need the  values of these
functions at small $z=\omega r$. The behavior in this limit is $\sim \Gamma(\nu)z^{-\nu}$, which at large $\ell$
will generate in the power expansion in $z$ huge coefficients due to the Gamma function $\Gamma(\nu)$.} The
calculation can then be organized as follows: ({\bf I}) determine the frequency $\bar \sigma$ beyond which the
probabilities $P_L(\sigma>\bar\sigma)$ become identical to those in the Minkowski space, and ({\bf II}) determine the
number $\bar L$ for which the cutoff is $\sigma_{cut}(\bar L) \simeq \bar \sigma$.  The effect of the curved
background on the detector's response will then be completely contained in the parameters
\bq
\sigma <\bar\sigma, \quad \ell\leq \bar L.
\eq
We plan to present an application of this procedure to a wormhole spacetime in a future paper.

\section*{Acknowledgements}

I thank my colleague Ailedi for inspiration and constant support.

\section*{Appendix}

We evaluate here (\ref{inteq}) in the limit $\varepsilon E \ll 1$. We mean by this  that we neglect all quantities of order
$\varepsilon$ or smaller. As mentioned in Sec.~3, the integral under the limit in (\ref{inteq}) is half of the excitation
probability at the time $\tau=\bar \eta$ for a sudden switch-on of the interaction at $\tau_0=0$ \cite{svai}.
We closely follow the calculation in Ref.~\cite{svai}. We split the integral in (\ref{inteq}) as
\bq
i(\bar \eta)=i_{I}(\bar \eta)-i_{II}(\bar \eta),
\nonumber
\eq
where $(\lambda\equiv\gamma \varepsilon E$)
\bq
 i_I(\bar \eta)= \int_{-\bar \eta}^{+\bar\eta} d\xi\,
\frac{\bar \eta}
{(\xi-i \lambda )^2}\,e^{-i \xi},
\label{i11}
\eq
and
\bq
i_{II}(\bar \eta)=\int_{-\bar \eta}^{+\bar\eta} d\xi\,
\frac{\vert \xi \vert}
{(\xi-i \lambda)^2}\,e^{-i \xi}.
\label{i2}
\eq

The integral (\ref{i11}) can be evaluated with a contour integral in the complex $\xi$ plane. We
consider that  $\xi$ varies over the entire real axis and close the contour in the semiplane Im $\xi>0$.
This leads to (the dependence on $\lambda\sim \varepsilon$ becomes irrelevant for a nonvanishing $\xi$)
\bq
i_I(\bar\eta)=
-\int_{-\infty}^{-\bar\eta}(\dots)
-\int_{\bar\eta}^\infty(\dots)\qquad\qquad\qquad \,\,\,\,\,
\nonumber
\\
=-2 \bar \eta \int_{\bar\eta}^\infty d\xi \frac{\cos\xi}{\xi^2}=
-2 \int_{1}^\infty du\frac{\cos(\bar \eta u)}{u^2}.
\eq
Since the cos factor becomes highly oscillatory for $\bar\eta\rightarrow \infty$ the conclusion is
\bq
\lim_{\bar \eta \rightarrow \infty} i_I(\bar\eta)=0.
\eq
In the second integral (\ref{i2})  it is sufficient to reorganize the integrand.
A few manipulations allow to write
\bq
i_{II}(\bar\eta)=\int_0^{\bar \eta} d\xi\, \xi \cos \xi
\left\{
\frac{1}{(\xi+i\lambda)^2}
+
\frac{1}{(\xi-i\lambda)^2}
\right\}
\nonumber
\\
= 2\int_0^{\bar \eta} d\xi \frac{\cos \xi -1}{\xi}
+j_{II}(\bar\eta),
\qquad\qquad\quad\,\,
\label{c1}
\eq
where (neglecting contributions of order $\lambda$ or smaller)
\bq
j_{II}(\bar\eta)=\int_0^{\bar \eta}
d\xi\, \xi \left\{\frac{1}{(\xi+i\lambda)^2}
+
\frac{1}{(\xi-i\lambda)^2}
\right\}=2(\ln \bar\eta -1) -2 \ln \lambda.
\label{c2}
\eq
Collecting the terms in (\ref{c1}) and (\ref{c2}) and paying attention to the numerical factor
in (\ref{inteq}) one arrives to (\ref{prolim}).  The constant $c$ is
\bq
c=\frac{1}{4\pi^2}
\left\{\lim_{\bar\eta \rightarrow \infty}
\left (
\ln \bar \eta +   \int_0^{\bar \eta} d\xi \frac{\cos \xi -1}{\xi}
\right)-1
\right\}
\nonumber
\\
=-\frac{1}{4\pi^2}(\gamma+1),
\qquad\qquad\qquad\qquad\qquad\qquad\,\,\,
\eq
where $\gamma$ is the Euler-Mascheroni constant.

\end{document}